# A review of machine learning approaches, challenges and prospects for computational tumor pathology

Liangrui Pan[1], Zhichao Feng[2], Shaoliang Peng[1]

[1]College of Computer Science and Electronic Engineering, Hunan University, Chang Sha, China

[2]Department of Radiology, Third Xiangya Hospital, Central South University, Chang Sha, Hunan, China

## Abstract

Computational pathology is part of precision oncology medicine. The integration of high-throughput data including genomics, transcriptomics, proteomics, metabolomics, pathomics, and radiomics into clinical practice improves cancer treatment plans, treatment cycles, and cure rates, and helps doctors open up innovative approaches to patient prognosis. In the past decade, rapid advances in artificial intelligence, chip design and manufacturing, and mobile computing have facilitated research in computational pathology and have the potential to provide better-integrated solutions for whole-slide images, multi-omics data, and clinical informatics. However, tumor computational pathology now brings some challenges to the application of tumour screening, diagnosis and prognosis in terms of data integration, hardware processing, network sharing bandwidth and machine learning technology. This review investigates image preprocessing methods in computational pathology from a pathological and technical perspective, machine learning-based methods, and applications of computational pathology in breast, colon, prostate, lung, and various tumour disease scenarios. Finally, the challenges and prospects of machine learning in computational pathology applications are discussed.

## 1. Introduction

According to the latest global cancer data released by the World Health Organization, the top five cancers with global incidence are breast cancer, lung cancer, colon cancer, prostate cancer, and stomach cancer. Among them, the top five cancers in cancer deaths are lung cancer, colon cancer, liver cancer, stomach cancer, and breast cancer [1]–[3]. Currently, the best method for diagnosing cancer has not been published. However, pathology images are the gold standard for diagnosing cancer. The traditional way of imaging histopathology is that pathologists rely on microscopes to examine slides, and the process remains manual. After the image on the slide is digitized, the pathologist can view the image remotely through a computer screen or mobile phone or analyze the image using image processing techniques [4]. However, the application of pathology images is still mainly based on subjective evaluation, lack of structured processing of image data, and insufficient mining of its hidden information. At present, the massive amount of pathological image data causes a serious burden on pathologists. Computational pathology can solve these two problems, provide intelligent diagnostic tools, and help to explore the connotation of potential disease mechanisms.

Initially, pathology images required the analysis of whole slide image (WSI) by clinicians and pathologists to guide cancer prognosis. However, for the observation of WSI with super large pixels, the human field of vision has certain limitations, introducing bias. The development of computational pathology can replace the manual selection of small regions of interest, automatically process the entire WSI, and perform quantitative analysis based on the information of different cells and tissues.

In recent years, image technology based on ML has been the focus of research, mainly focusing on image segmentation, classification, and quantification. ML has enabled tremendous progress in image analysis, generating powerful algorithms with fewer iterations on datasets than manually observing WSIs. As a branch of ML, deep learning (DL) has been widely used to process histopathology images. Its feature extraction method can extend from the local view to the global view, analyze the image, find the view of interest, and display it in the form of a heat map. However, the interpretability of DL algorithms still needs to be improved, and some new techniques compensate for some of the shortcomings of DL to some extent, such as transformer [5].

Computational pathology is most important to help pathologists and doctors reduce the probability of misdiagnosis. Currently,

a large number of global competitions for histopathology images are being promoted, which contributes to the development of computational pathology. In the future, computational pathology has the potential to change the core functions of traditional pathology, not just the growing subfields of digital pathology, molecular pathology, and pathology informatics [6]–[8] . Second, computational pathology increases the reliability and accuracy of diagnosis through global collaboration to reduce physician workload and patient treatment costs. Third, as technology advances, computational pathology will gradually become part of personalized precision medicine. As shown in Fig. 1, machine learning methods for processing histopathology images require three steps: data collection and pre-processing; machine learning methods; segmentation, classification, and quantification of histological images.

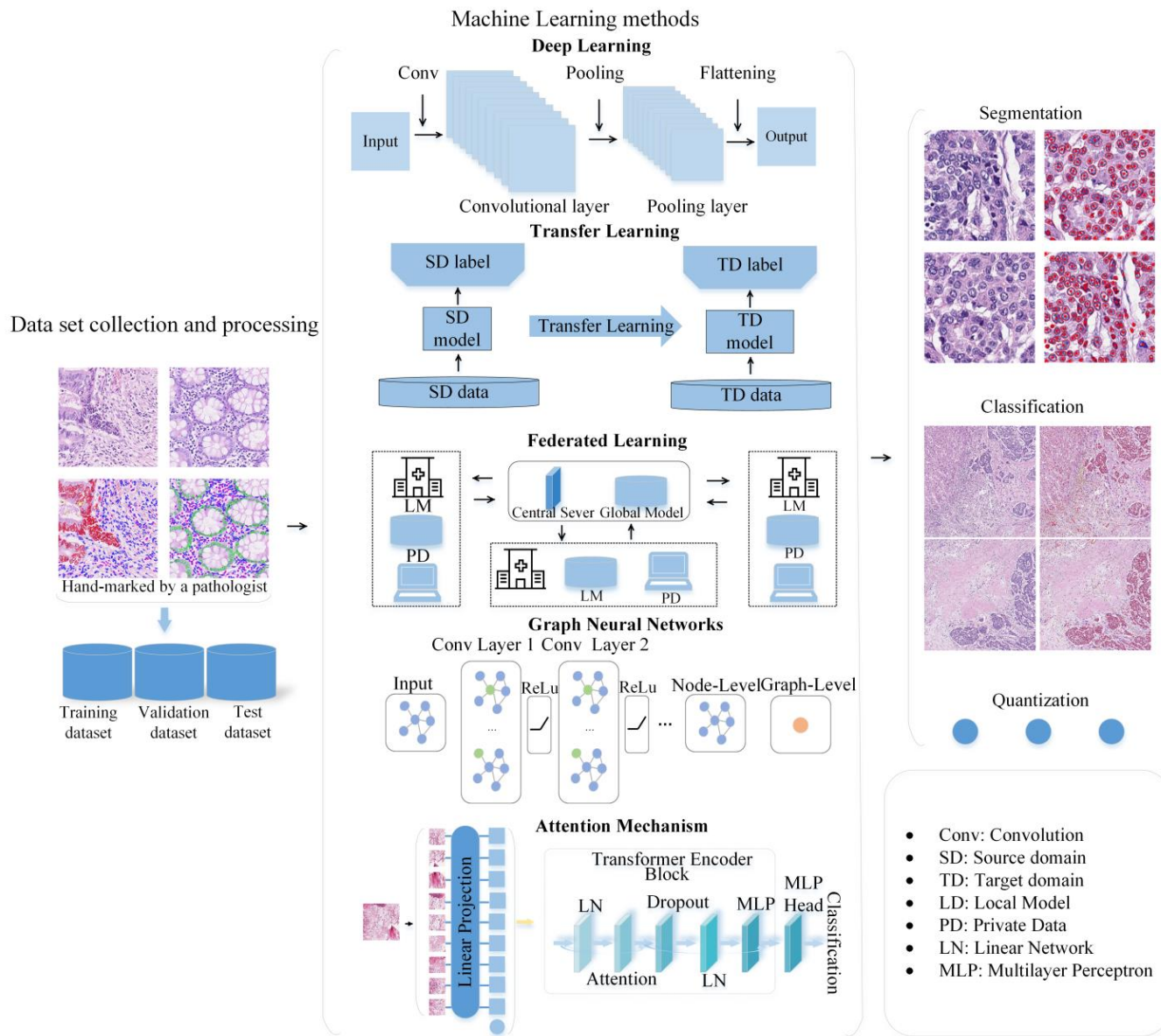

Fig .1. Flow chart of computational pathology processing histopathology images.

We conduct an extensive review of papers containing keywords such as (Pathology or Histopathology or Computational Pathology), (classification or segmentation or histological) and (Breast or Colon or Lung) through Google Scholar, PubMed and arXiv using extensive keyword searches or Prostate Cancer) In addition, we also searched the conference proceedings of MICCAI, ISBI, MIDL, SPIE and EMBC based on the title, abstract and keywords of the paper. Given the rapid growth in the number of relevant articles on this research topic, we reviewed 201 articles published from 2018 to April 2022 and categorised and summarised 126 articles. Fig. 2 shows that this review counts the articles involved according to year. The rest of the article is organized as follows: Section II introduces data preprocessing; Section III introduces machine learning methods for processing histopathology in computational pathology; Section IV introduces computational pathology in breast, colon, and prostate Application of classification, filamentous classification detection, segmentation and quantification on cancers such as cancer and lung cancer; Sections V and VI introduce the four major challenges and prospects of machine learning in computational pathology applications, respectively.

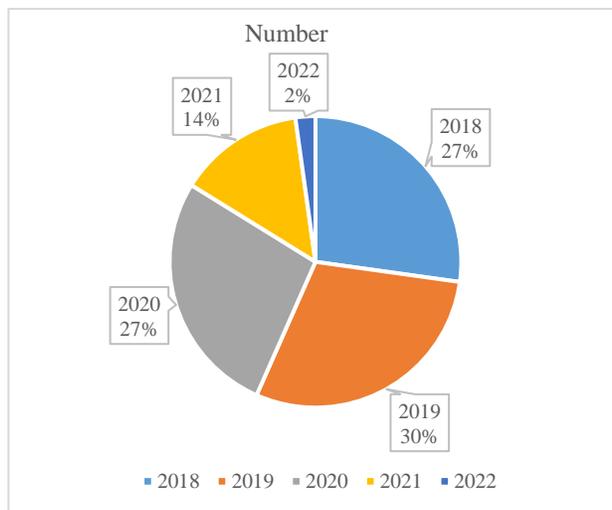

Fig. 2. Counts relevant literature in different years.

## 2. Computational pathology data

**2.1. Data preparation**

Pathology images contain rich and detailed information from the tissue to the cellular level. A typical full-slice image is about 50 000 × 50 000 pixels, equivalent to a single whole slice containing 10 GB of uncompressed pixel data. At present, advanced pathological scanning equipment such as Leica and Philips in the industry can scan a 2×3-inch glass slide with a pixel resolution of 0.25 μm, and the obtained digital image of the WSI can reach 200,000×300,000 pixels. Pixel raw data up to 240 GB [8], [9]. Taking Memorial Sloan Kettering Cancer Center (MSKCC) as an example, about 1 million new pathological slides are generated every year, and the original data of 1~24 EB (1018 magnitude) will be generated after all digitization [10]. This poses a huge challenge for data storage and loading. Furthermore, the memory associated with the central processing unit (CPU) or graphics processing unit (GPU) is usually limited. Therefore, pathology images should first be cut into small pieces and/or resized to fit CPU or GPU memory if necessary [11]. 256 × 256 or 512 × 512 pixels are the most common image slice sizes in experiments.

**2.2. Data preprocessing**

In order to speed up the model's reading speed of data and the generality of training, it is necessary to perform image preprocessing on images before training. In the pre-processing stage, experiments usually require a normalisation operation on the image and enhancement of the WSI features [12]. In practice, many algorithms work best after normalizing and whitening the data. Normalization does not change the image information but only changes the pixels from 0-255 to the range of 0 to 1 to speed up the convergence of the training network; the specific function of normalization is to summarize the statistical distribution of the unified sample. Normalized between 0-1 is a statistical probability distribution. Common normalization methods are ① simple scaling; ② sample-by-sample mean reduction (also known as removing the DC component); ③ feature normalization (making all features in the dataset have zero mean and unit variance) [13], [14].

Data augmentation, also known as data augmentation, aims to increase the quantity and diversity of limited data, aiming to extract more useful information from limited data and generate value equivalent to more data. In a broad sense, data augmentation can be divided into data warping and data oversampling to generate new data. Due to the simplicity of operation and the much lower demand for data volume, the early applications of data augmentation were mostly data deformation methods. The basic image transformation operations belong to data deformation augmentation methods for image data, and the earliest work applied to DL can be seen in LeNet-5 for the affine transformation of images [15]. With the development of convolutional neural networks, various classical network models have more or less adopted data-morphing augmentation methods in their image classification tasks to prevent overfitting. For example, AlexNet uses cropping, horizontal mirroring, and colour enhancement based on principal component analysis (PCA) to augment the training data; the VGG network uses multi-scale scaling and cropping to augment the data; GoogLeNet takes the cropping plus mirroring approach of AlexNet and extends it further by augmenting the data at test time, augmenting an image to 144 samples and averaging the Softmax probabilities of all samples to obtain the final classification results; similarly, in the later ResNet and densely connected convolution, the results are augmented with the Softmax probabilities of all samples [16]–[20]. As the effectiveness and necessity of data augmentation have been verified in more and more experiments, many researchers have begun to explore new data augmentation methods.

Pathology images are very different from other medical images due to different staining conditions and slide thicknesses, so DL algorithms need to adapt to images from different domains [11]. Normalising pathology images to a uniform scale is a common solution. There are several pathology image normalisation methods that researchers have applied. However, they are usually time-consuming and may reduce some of the intrinsic information. For example, hematoxylin and eosin-stained pathological images of renal cell carcinoma are usually divided into eosinophilic and basophilic subtypes, in which the chromatin and nucleic acids within the nucleus are readily stained purple-blue by hematoxylin and the cytoplasm and extracellular matrix are stained red [11]. A relatively simple solution in a DL environment is using a colour enhancement to simulate the actual differences. By adding a random mean to each RGB channel of each image and multiplying by a random variation, the sample size is greatly increased. As a result, DL models can learn to ignore the systematic biases that arise during the production of pathology sections to obtain models with robustness. When it comes to colour enhancement, we should carefully choose the range, mean, and variance of the enhancement parameters to reduce the inherent characteristics of distorted images [11]. Other colour enhancement methods include the addition of Gaussian noise, the introduction of pretzel noise and blurring [21].

At present, the most common data augmentation methods are summed up as geometric transformation, gamut transformation, definition transformation and noise injection. Geometric variation is to generate new samples through rotation, mirror image, translation, clipping, scaling and distortion [22], [23]. Fig. 3 visually depicts 20 schemes for data augmentation. These protocols are commonly used methods for preprocessing histological images. Data augmentation based on gamut transform is essential to enhance the robustness of the model under different illumination conditions by adding various illumination brightness deviations to the data set. Definition transform, also known as "nuclear filter", is a new sample generation method to change the visual definition of images. Finally, noise injection is used to introduce redundancy and interference information to the dataset by artificially applying noise interference to the images; simulating images of different imaging qualities; enhancing the model's ability to filter noise interference and redundant information; improving the model's ability to recognise images of different qualities.

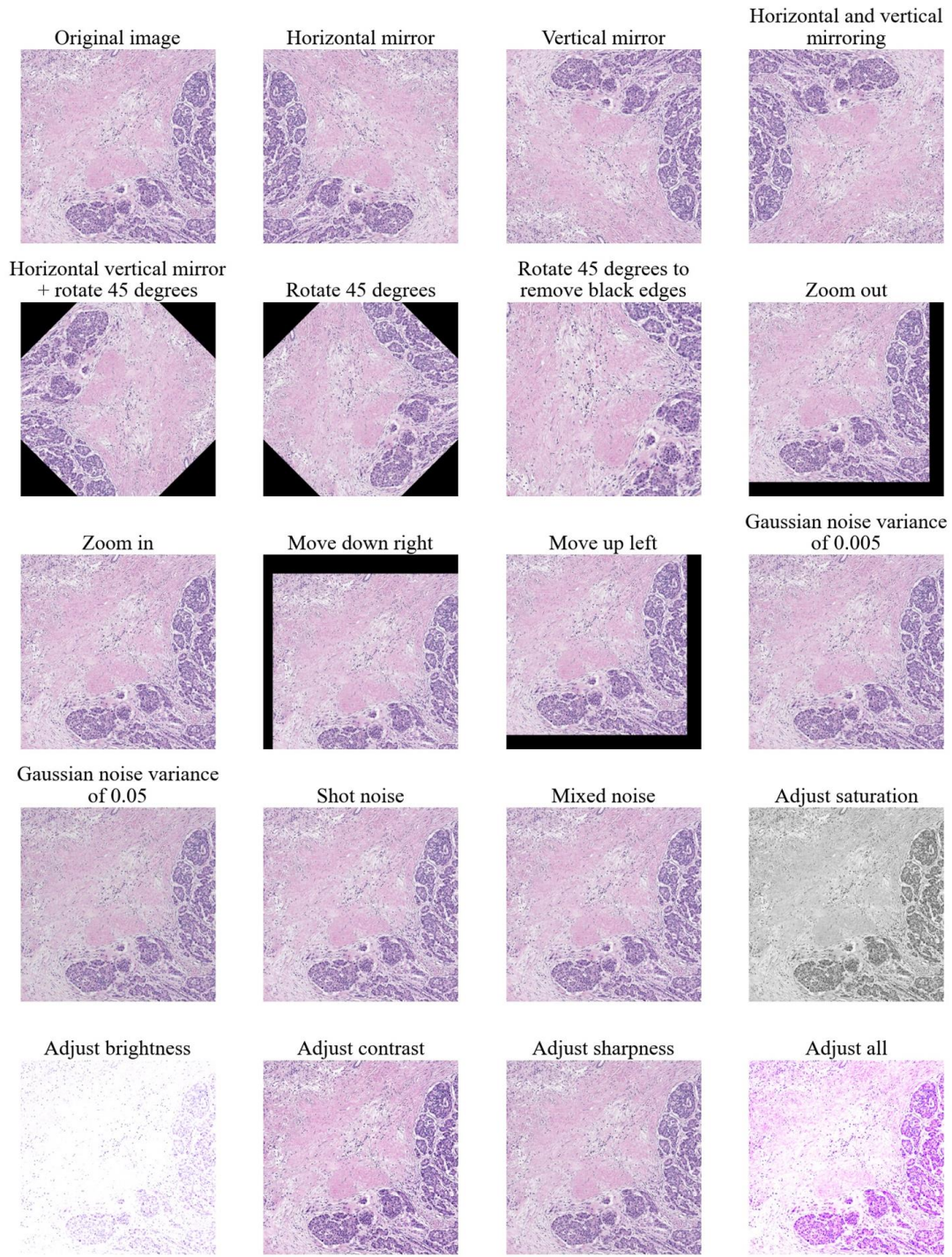

Fig. 3. Visualizes some data augmentation schemes, which are horizontal mirror, vertical mirror, horizontal and vertical mirroring, rotate 45 degrees, horizontal vertical mirror + rotate 45 degrees, rotate 45 degrees to remove black edges, zoom out, zoom in, move down right, move up left, gaussian noise variance of 0.005, gaussian noise variance of 0.05, shot noise, mixed noise, adjust saturation, adjust brightness, adjust contrast, adjust sharpness and adjust all, respectively.

## 3. Machine learning based methods

We summarized the technology from the popular research areas of machine learning. Fig. 2 shows that the main machine learning directions in tumor computational pathology are DL, transfer learning, federation learning, graph convolution, and attention mechanisms. However, there are often cross-cutting relationships among these techniques, such as deep transfer learning, a combination of DL and transfer learning. Therefore, this review analyzes and summarizes these techniques as an independent research direction.

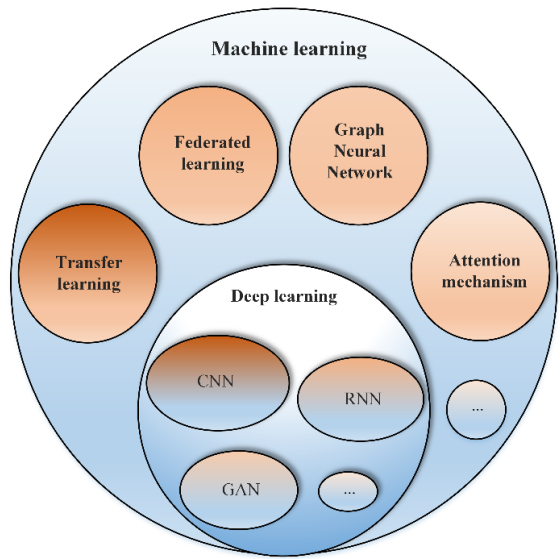

Fig .4 The directions and relationships of different researches in the field of machine learning.

## 3.1. Deep learning

DL is one of the hot research areas in ML, where deep (multi-layer) network architectures are used to map the connections between input or observed features and outputs. This deep architecture makes DL particularly suitable for dealing with problems with many variables. At the same time, the features generated by DL can be regarded as an integral part of the learning algorithm rather than feature generation as a simple step. DL can represent more and more abstract concepts and patterns level by level and train algorithms end-to-end. It has been proved that DL has succeeded in computer vision and natural language processing [24], [25]. Thanks to these excellent algorithms, cheap computing power, and the sharing mechanism of big data, some software and tools based on DL algorithms have begun to be implemented. Subsection 3.1 will introduce the three branches of DL: convolutional neural networks, recurrent neural networks, and generative adversarial networks.

### 3.1.1. Convolutional Neural Network

Convolutional Neural Network (CNN) is mainly composed of the input layer, convolutional layer, pooling layer, ReLU layer and fully connected layer [26]. A complete CNN can be constructed through the superposition of these layers. Neurons in a CNN are arranged in width, height, and depth, where height and depth are the length and width of a histological slice. The convolutional layer is the core layer for building a CNN neural network, and it undertakes most of the computation in the algorithm. The convolution operation is essentially a dot product between the filter and a local region of the input data. The convolution layer uses the convolution operation to turn the result of forwarding propagation of the convolution layer into a huge matrix multiplication carrying pathological features. The convolved part, the spatial size of the convolution kernel connection, is called the receptive field. The size of this connection is always equal to the depth of the input. Weight sharing in convolutional layers is used to control the number of parameters. However, neurons of different depths do not share the same weights during backpropagation, so only one set of weights can be updated. The pooling layer is usually placed after the convolutional layer to gradually reduce the spatial size of the data volume. In this way, the number of parameters in the network can be reduced, the computational resource consumption can be reduced, and overfitting can be effectively controlled. Common max-pooling layers operate independently on each depth slice of the input data volume, changing its spatial size [27]. Besides max pooling, the pooling unit can also use other functions, such as average pooling or L2-norm pooling. When forward-propagating through the pooling layer, the index of the largest element in the pool is usually recorded (sometimes called switches) so that gradient routing is efficient when back-propagating. The ReLU layer generally selects the ReLU function as the activation function. Compared with the sigmoid function, the computational cost is much lower, and secondly, it can alleviate the problem of gradient disappearance. The primary function of the fully connected layer is to map the learned "distributed feature representation" to the sample label space. Its essence is to transform from one feature space to another, namely classification [28].

CNN uses the convolution kernel in the convolution layer to extract the features of the image. The multi-channel feature enables the features to share weights in the forward transmission process, which reduces the complexity of the model and the number of weights. The input of WSI as a whole avoids complex feature extraction and data reconstruction. The normalization operation accelerates the training of the model so that the model reaches convergence. Among them, nonlinear characteristics can extract more abstract features in feature extraction. The local perception feature of CNN can regard WSI as an object to be processed, and the convolution kernel only calculates the local cell features and other details of WSI each time the convolution operation is performed. After feature extraction, this local information is summarized and sorted at the end of the model, and the results are output. For example, the common LeNet, AlexNet, VGG, GoogleNet, ResNet, Densenet, Senet, BAM, mobilenet, etc. are all part of CNN [16]–[20]. Now, these models are only used as a module for feature extraction in histopathological image analysis, combined with other functional modules to form an end-to-end image analysis framework.

### 3.1.2. Recurrent Neural Network

Unlike CNNs and FCNs, which are limited to the analysis of data from one individual time point, recurrent neural networks (RNN) can store inputs over different time intervals or time points in order to process them sequentially and learn from (often several million) discrete earlier steps [29], [30]. RNN hidden nodes form memory in a circular structure. The state of the hidden layer at each time depends on its past state. This structure enables RNN to save, remember and process long-term past complex signals. However, sometimes, we need to use the latest information to deal with the current task. Because the colour WSI image is three channels, for each WSI, we can regard it as a matrix. Like a sentence in natural language processing, RNN regards each row of WSI as a vector and processes WSI from top to bottom or from left to right in the order of row arrangement. The input gate of RNN will process the WSI image, and the forgetting gate will extract the effective map image features in WSI. The filtered useless features will not need to continue to transmit, such as interstitial or noise information. The output gate will determine which knowledge needs to be transferred to the next series. Therefore, we know that the longer the RNN sequence, the better the effect of feature extraction. Its end will collect the weights from all previous features and output the cell category of WSI.

However, due to the problems of gradient dispersion and gradient explosion in the original RNN, part of the information on the WSI will be lost during feature extraction. Based on the RNN mechanism, a special structure is proposed as Long Short Term Memory (LSTM) [31]. The first step of an LSTM is to pass the input through a forget gate containing a sigmoid function to determine what information to forget from the input. The second part needs to store some new information in the state unit. Finally, update the old cell state to the new cell state and determine the cell's output. Besides LSTM, some RNN-based networks have also been proposed, such as bidirectional recurrent neural networks, deep RNN, etc. [31]. Second, the complementarity of CNN and RNN helps the model to improve sensitivity and accuracy to a certain extent. RNN is often used as the output module of image analysis results because it can effectively combine context information for accurate reasoning.

### 3.1.3. Generative Adversarial Networks

Besides CNNs and RNNs, Generative Adversarial Networks (GANs) are also showing increasing promise in the field of computer vision. GAN mainly consists of two parts, which are a generator (generating network) and a discriminator (discriminating network) [32]. The role of the generator is to learn the features of the real WSI. Under the guidance of the discriminator, the random noise distribution is fitted to the distribution of the real image as closely as possible, thus allowing the generator to generate similar data with real image characteristics to fool the discriminator. The discriminator is responsible for distinguishing whether the input data is real or fake data generated by the generator and feeds it back to the generator. The two networks are alternately trained to achieve a dynamic balance of capabilities: the data generated by the generative network can be faked, while the discriminator cannot identify real and fake images, that is, the probability that a given image is predicted to be true is close to 0.5. In the WSI classification, we will replace the last layer of the discriminator with a softmax layer and output the probability of the cell. The advantages of GANs are as follows: they can better model data distributions, do not require iterative sampling using Markov chains, do not require inference during learning, do not have complex variational lower bounds, and bypass the difficult problem of approximating the computation of tricky probabilities [32].

In contrast to non-linear networks, GANs do not require the latent variables of the generator input to have any specific dimension or require the generator to be invertible. However, some shortcomings of GAN have to be mentioned, the interpretability of the model is poor, it is not easy to converge during training, or the gradient disappears and the mode collapses. These issues are critical to medical imaging, and unconvincing models will be questioned among doctors and patients. Nevertheless, with the continuous improvement of the GAN model, it is believed that there will be more suitable GANs to deal with WSI in the future.

### 3.2. Transfer learning

Similar to DL, transfer learning (TL) is another research hotspot in machine learning. At present, transfer learning has been widely used in histopathology images. It learns the features of the new WSI through the existing WSI features, and the core is to find the similarity between the existing WSI and the new WSI. Since it is too expensive to learn from scratch directly on a new dataset of histopathology images, we will use the existing relevant features to assist the model to learn new WSIs features as quickly as possible. Specifically, in transfer learning, we call the existing empirical knowledge of the model as the source domain and the new features to be learned by the model as the target domain. Transfer learning studies how to transfer experience from the source domain to the target domain. In particular, in machine learning, transfer learning studies how to apply the existing experience in the model to a new and different research field, but there is a certain correlation between the new and the old field. When traditional machine learning is dealing with tasks such as data distribution, dimensions, and changes in the output of the model, the model is not flexible enough, and the results are not good enough, and transfer learning improves these situations. Under the condition of data distribution, feature dimension and model output changes, the experience in the source domain is efficiently used to model the target domain better. In addition, in the absence of calibration data, transfer learning can use calibrated data in related fields to complete data calibration.

According to the learning method, transfer learning can be divided into the sample-based transfer, feature-based transfer, model-based transfer, and relation-based transfer [33], [34]. In theory, data between any domains can be modeled using transfer learning. However, if there is not enough similarity between the source and target domains, the modelling results will not be satisfactory and a so-called negative migration situation will occur. For the histopathology image dataset, we can first pre-train the model with some of the data and put the generated pre-trained model into the new histopathology dataset for secondary training to obtain the final training model. In this process, we only fine-tune the pretrained model. The performance of the model will be significantly improved, and its convergence will be better. There is a close relationship between transfer learning and CNN. As a feature extraction module of transfer learning, CNN will accelerate the transfer learning strategy to play a role in computational pathology. This approach is also known as deep transfer learning [35].

### 3.3. Federated learning

Unlike DL and TL, federated learning (FL) is a distributed ML technology with privacy protection and secure encryption or a ML framework. It is designed to allow decentralized participants to collaborate on ML model training without disclosing private data to other participants [36]. FL can be divided into horizontal FL, vertical FL, and federated TL according to the different ways of contributing data sets [36]–[38]. The essence of horizontal FL is the union of samples, and it is suitable for scenarios where participants have the same business format but reach different customers, that is, when there is much overlap in features and little overlap in users. The steps of horizontal FL are: ① Different users download the latest model from the cloud shared server; ② Each participating user uses local data to train the downloaded model, and encrypts the gradient and uploads it to the shared server, and the server aggregates the gradients obtained by each user to update the model parameters; ③ The server generates a shared model and provides it to each user for download; ④ Each user uses the local data to update their model again [39]. The essence of vertical federated learning is the combination of features, which is suitable for scenarios where users overlap a lot and features overlap less. The steps of vertical FL are: ① The public server sends the public key to the client to encrypt the data to be transmitted. ② Different users calculate the intermediate results of their features respectively and encrypt the interaction to obtain their respective gradients and losses; ③ Each user calculates their encrypted gradients and adds masks to interact with other users.

④ Other users decode the gradient and loss and then pass it to each user to remove the mask and update the model. FL learning can be considered when there is little feature and sample overlap between participants. Its steps are similar to longitudinal FL, but the intermediate pass results are different (in fact, the intermediate pass results are different for each model). The advantages of FL can strengthen academic cooperation and exchanges between different countries and regions and promote the technological progress and development of computational pathology.

WSI data from different institutions can directly train diagnostic models locally. At the same time, all institutions can upload WSI data and local models to a shared server, and different institutions can download models from each other for use. However, the prior knowledge obtained by the shared model is insufficient. The common method is to download a large amount of data from different institutions from the shared service and then retrain the local pre-trained model to expand the prior knowledge of the model. First, the institution will upload the locally trained final diagnosis model to the cloud server for model sharing. Second, other institutions can download the diagnostic model on the cloud server or upload the WSI to the cloud for online analysis of histopathology images. The method efficiently utilizes internet and computing resources, coordinates data and shares diagnostic models.

### 3.4. Graph Neural Network

Graph Neural Networks (GNNs) refer to methods in applying DL to structured data. Since the structure of graphs is not regular, traditional deep networks are not easy to generalize to graph-structured data. Therefore, GNN can be seen as a learning process about the features of the graph. For tasks that focus on nodes, the GNN model aims to learn representative features of each node, which will help the subsequent processing of such tasks. For graph-focused tasks, the goal of GNN models is to learn representative features of the entire graph, and learning node features are usually only an intermediate step in it. The process of learning node features usually uses both node input features (attributes) and graph structure. The most commonly used algorithms in GNNs are graph CNNs, graph attention networks, graph autoencoders, graph generative networks and graph spatiotemporal networks [34] [40]. One of the largest application areas of GNNs in computer vision.

The graph neural network is combined with convolution to form a graph convolutional neural network. The essence of the convolution operation in CNN is to perform a weighted summation of pixels in a range, which helps extract the spatial features of WSI. However, graph convolution is a convolution operation based on the graph's topological structure. Each pixel is regarded as a point constituting the graph. The feature is actually a weighted average of all surrounding pixels after propagating the feature value to the center point. Similar to CNN, graph convolution also adopts weight sharing. The weight of the kernel in graph convolution is assigned according to the corresponding position, and the weight in graph convolution is usually a set. When calculating the aggregated eigenvalues of a node, all points participating in the aggregation are allocated into multiple different subsets according to a certain rule. The nodes in the same subset adopt the same weight to realize weight sharing.

### 3.5. Attention mechanism

The attention mechanism (AM) borrows from the way of thinking of human attention. By quickly scanning the WSI, the model needs to obtain the target area to focus on, which is also referred to as the focus of attention. More attentional resources are then devoted to this region to obtain more detailed information about the desired target of attention, thus suppressing other useless information [41]. Therefore, human visual AM greatly improves computer visual information processing efficiency and accuracy. Due to the great success of AM in natural language processing, AM is gradually used in the field of computer vision. Suppose most of the current methods are described in abstract terms. In that case, the specific computation of the Attention mechanism can be summarised in two processes: the first is the computation of the attention weighting coefficients based on the Query and Key from the different images, and the second is the weighted sum of the Values from the images based on the weighting coefficients. The self-attention mechanism (SAM) is a variant of the AM that reduces the reliance on external information. Given that the Query, Key and Value of the SAM come from the same image sample, it is better at capturing the internal correlation of data or features. The application of the self-attention mechanism in computer vision mainly solves the problem of feature extraction and classification of images through the similarity coefficient between images [42], [43]. Compared with CNN and RNN, it has the advantages of smaller complexity, fewer parameters, faster speed and better effect.

Compared with CNN and RNN, the attention mechanism has the advantages of smaller complexity, fewer parameters, faster speed and better effect. First, the attention mechanism can perform feature learning on the entire WSI for cell features on pathology images, and slices of different sizes can enable the attention mechanism to obtain more delicate features. Furthermore, the position information's embedding makes the image information's traceability basis. The pathological image will not misdiagnose the content of the image because of the feature similarity of the local area. Second, focusing the limited attention on the pathologist's region of interest will facilitate feature extraction, speed up the training of the diagnostic model, and reduce the parameters of the model. At present, the popular Transformer is being widely used in image recognition and detection [5]. First, it focuses attention on the sensitive areas of WSI to improve the efficiency of feature extraction. Secondly, the fine-grained feature extraction mode enables the model to obtain more accurate prior knowledge [44].

## 4. Applications of Computational Pathology

The main clinical uses of computational pathology include lesion identification, disease diagnosis, differential diagnosis, grading or subtype diagnosis, prognosis prediction, efficacy assessment, and genomic prediction. Accurate early detection reports show promise for cancer diagnosis. Machine learning-assisted systems accomplish classification and segmentation tasks with unprecedented accuracy and lay the groundwork for the deployment of computational pathology in the field of cancer research. Machine learning approaches complement medical-related expertise and support pathologists and oncologists. This section will do a comprehensive literature survey from three aspects: segmentation, classification and quantification.

**4.1. Segmentation and detection**

Histopathology images contain much biological information, such as cell size, nuclear morphology, mitotic status and interstitial fluid distribution. Unfortunately, human vision is challenging to capture quickly, and the quality of acquired image information is relatively poor. Therefore, the segmentation of WSI is a significant step in analyzing histopathology. When we divide the cells, mark the cells or mark the mitosis, we know the cell distribution information and cell number information on this WSI. Based on this valuable information, pathologists or clinicians will further analyze the patient's condition and assist the clinic in formulating a diagnosis plan.

**4.1.1. Pathology image segmentation**

Glands are the main basis for the diagnosis of colon cancer, and accurate segmentation from histopathology images is the premise for judging the grade of colon cancer. Colon cancer is more concerned about the performance on the WSI. A Segnet-based gland segmentation method was the first to segment A and B parts on the Warwick-QU dataset with an accuracy of 0.882 and 0.8636 [45]. Sparsely annotated histopathology data is difficult for pathology images, and DL seems to be a promising solution [46], [47]. Some researchers rely on large-scale computing resources to train DL models and perform patch-level classification on the AiCOLO colon cancer dataset, and the results validate the strong advantages of the models [48]. Hover-Net is a new CNN-based model that can effectively instantiate and segment colorectal adenocarcinoma image patches [49]. The improved UNet and Segnet models based on DL can make up for the lack of rich annotations in WSI [50]. Second, the DCIA framework exploits the features generated by DenseNet to explore optimal representations for modelling instance segmentation [51]. The framework also incorporates morphological methods and convolutional conditional random fields (ConvCRF) to refine confidence maps for more accurate segmentation results [51]. The lack of pixel-level annotation data can lead to computational pathology techniques accuracy and reliability. Other researchers collected histopathology images of 19 patients with liver metastatic colon cancer and asked two pathologists and one clinician to Perform image annotation [52]. They developed SVM, kNN, U-Net, U-Net++ and DeepLabv3 classifiers to instantiate and segment public dataset images. DL classifiers outperformed SVM and kNN by 14% in micro-balanced accuracy, 15% in micro-F1 scores, and 26% in micro-precision [52]. Collaboration across multiple institutions enables colon cancer digital pathology images to collect and share to obtain histopathology images of colon cancer patients. The vast amount of data has improved the accuracy of segmentation of the model and the accuracy of cancer prognosis. Using the tissue appearance of colon glands to delineate gland boundaries has been shown to segment internal glandular structures, including epithelial cells, nuclei and lumen, and provides the basis for segmentation across multiple cancers [53].

In order to accurately delineate the boundaries of the glands, a colour map-enhanced image sharpening method can obtain an enhanced image in which all key elements are easily detected from the original histopathological image [54]. This technique significantly improves the interpretability of content information in histopathology images. Second, a novel minimal information loss dilation network (MILD-Net) was also proposed for accurate gland instance segmentation in colon histology images [55]. A fully convolutional neural network reintroduces the original image at multiple points to cope with information loss caused by max pooling. It demonstrates the effectiveness of the proposed method by performing gland instance segmentation on two other datasets containing whole slide images. Scalability [55]. Fig. 5 shows the visual effect of the MILD-Net model segmenting glands and cavities. In order to improve the segmentation efficiency of the model, the HistoSegNet model uses the HTT-annotated Atlas of Digital Pathology (ADP) database to train the WSI semantic segmentation model, and the method can be re-applied across datasets [56]. A discriminative error prediction network for semi-supervised colon gland segmentation can effectively utilize images with intra- and inter-class errors to obtain reliable segmentation models [57]. The network proposes a more reliable self-training double error correction method for unlabeled parts that can effectively display glandular regions [57]. DL is still the preferred method for segmentation tasks. Attention-Guided Deep Atrous-Residual U-Net adds attention mechanism and atrous-residual unit based on the improved U-Net architecture, which can extract more features and multi-level feature representation , the problem caused by the gradient decrease of the link model [58]. The method is extensively experimentally validated and validated on two public datasets and one private dataset and achieves state-of-the-art results. Segmentation Model After three years of development, SegCenterNet2 is currently the state-of-the-art model for colon cancer diagnosis and has been shown to be superior to Mask-CNN in the CoNIC challenge dataset [59].

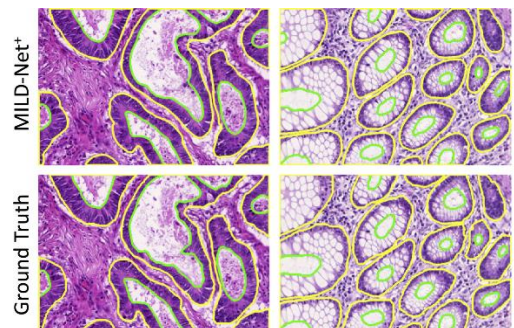

Fig. 5. Visual effects of the MILD-Net model segmenting glands and cavities. Among them, the first row shows the output of the MILD-Net model. The second row shows the pathologist annotations. The yellow outline shows the outline of the glandular border, and the green outline shows the outline of the lumen border [55].

Pathologists want to process WSI for breast cancer prognosis by computer vision [60], [60], [61]. First, the experiment uses K-means clustering to generate an initial segmented image and then applies the watershed segmentation algorithm. Experiments demonstrate that the method effectively detects and classifies breast cancer from histological images with an accuracy of 70% [62]. According to the different magnifications of histological images, image segmentation models based on particle swarm optimizer (PSO) simplify the image representation and segmentation task [63]. The stochastic fractal search (SFS) algorithm and minimum cross-entropy as the objective function achieved the best performance for the segmentation of breast histology images [64]. Although CNN-based U-Net has achieved good results on histological image segmentation tasks, some researchers prefer to use DL to develop algorithms for breast cancer cell nucleus segmentation [65] [61]. An efficient encoder-decoder architecture based on the DL model Separable Convolutional Pyramid Pooling Network (SCPP-Net) can efficiently segment breast histopathology images [66]. A Deep Residual Neural Network (DeepRNNetSeg) model learns high-level discriminative features of kernels from pixel intensities and generates probability maps [67]. The model's F1 score improved to 0.8513, and the average accuracy reached 86.87%.To achieve interpretable diagnostic results, the DeepBatch framework generated by combining preprocessing, ROI detection, ROI sampling, and region segmentation modules based on a CNN cascaded framework can assist pathologists in providing interpretable and accurate predictions for patients [68]. To reduce annotation time, a deep multi-magnification network (DMMN) is used to train using partially annotated images, followed by accurate multi-class tissue segmentation of the entire WSI [69]. T Avoiding the dependence of deep learning on big data, a number of pathologists and physicians recruited 25 participants to describe an active contour model for adaptive ellipse fitting (CoNNACaeF) using digital slide archiving and segmentation on three breast histopathology image datasets [70]. However, insufficient data will lead to poor segmentation of the model. Some researchers have developed a method to semi-automatically collect a large number of annotations to complement the largest dataset available today, PanNuke [71].

The classification and grading of prostate cancer is often based on the characteristics of pathology images. Localized prostate cancer adjudication sometimes requires co-registration of 18 F-choline PET/CT with T2-weighted 3D sequences and semi-automated segmentation procedures to define tumour volume [72]. With the development of 3D technology, comprehensive 3D imaging through the OTLS platform can display 3D features of glandular morphology. The 3D watershed algorithm can then segment the glands, filter out false positive areas, and highlight sensitive areas [73]. The U-Net model is the dominant method for histological image segmentation [74], [75]. It can use colour deconvolution to preprocess a subset of immunohistochemical (IHC) slides to segment epithelial structures in IHC [76]. Reducing the channel dimension and adding an extra downsampling step in the U-Net network can improve the accuracy of segmenting histological images by 3.6% [77]. Validation on external data demonstrates the segmentation effect of the U-Net model, but only as part of a fully automated prostate cancer segmentation. Convolutional neural networks reduce the complex pipeline of histopathology images, and FCN-8, two SegNet variants, and multi-scale U-Net showed high accuracy in segmenting high- and low-grade tumors, respectively [78], [79]. The feature extraction effect of the residual network and the U-net model achieved an average Dice Index of 0.77 in the test subset [80]. Mathematical morphology, an image analysis technique, can successfully define the borders of prostate cancer [81]. In order to improve the segmentation accuracy of the model, the model first uses the MS (mean shift) algorithm for coarse segmentation and then uses the wavelet recorder to perform fine segmentation of the tissue glands. Fully labelled data is at a premium, and simple morphometric techniques can generate weakly labelled data to help train convolutional neural networks for pixel segmentation of stromal, epithelial, and luminal (SEL) regions [82]. DL remains the method of choice for processing histopathology images [83]. The current state-of-the-art convolutional network architecture (NASNetLarge) based on DL models needs to rely on large, high-quality annotated training datasets, and its models achieve over 98% when augmenting the policy [84]. Second, the model can directly perform Gleason grading on the prostate with an accuracy of human-level performance. A hybrid DL approach based on a softmax-driven active contour model can accurately detect and delineate the contour of the prostate [85]. Comparing the results of the automatic segmentation with manual annotation and 7 advanced techniques, the method can delineate the prostate of unknown histopathology images with a Dice index of 90.16%. The Gland Context Network (GCN) enhances any glandular feature and characterizes prostate samples with context. Its results show that GCN can help define effective prostate grading methods and also obtain state-of-the-art classification performance [86].

DL performance is limited by the scarcity of large-scale fully annotated datasets. An end-to-end trainable deep neural network based on the feature union of two branches extracts high-level feature maps from low-level feature maps through independent branches, respectively, and generates segmentation models [87]. This method uses transfer learning to reduce the impact of insufficient training data and conducts large-scale ablation experiments on public benchmark datasets, proving the method's superiority. Augmenting a model trained on a limited dataset with external weakly labelled data will improve the segmentation accuracy of the model. The expectation-maximization-based approach with approximate prior distribution constraints on the semi-supervised method trained on 135 fully annotated and 1800 weakly annotated patches achieves an average Jaccard index of 49.5% on the independent test set, better than only the fully annotated dataset. 14% higher than the initial model trained on [88]. An unsupervised tissue cluster-level map cut (TisCut) method divides histological images into meaningful intervals (e.g. tumour or non-tumour) to assist downstream supervised models of annotated histological images [89]. Compared with the U-Net model, the TisCut method achieves a Jaccard index of 85% in dividing histological images.

Pathology images are also required to diagnose other cancers, such as lung cancer, liver cancer, and gastric cancer. A multi-scale receptive field deep learning (Deep-Hipo) model extracts patches of the same size at both high and low magnifications and captures complex morphological patterns in the receptive field of the whole slide image [90]. It outperforms state-of-the-art DL algorithms in segmenting gastric cancer histological images. However, the scarcity of annotations severely hinders the study of diagnostic models. Weakly supervised frameworks rely on standard clinical annotations, leveraging multi-instance protocols to train models to segment whole slide images of different cancers in the TCGA database [91]. The NucleiSegNet model was proposed to facilitate the segmentation of nuclei in histopathology images of liver cancer [92]. The attention decoder block of this method uses a novel attention mechanism for efficient object localization and improves the performance of the proposed architecture by reducing false positives [92]. Finally, hybrid feature segmentation can achieve the associated segmentation objective-assisted model segmentation task objective. A multi-level thresholding image segmentation method based on the Rényi two-dimensional entropy and culture algorithm (2DR) can segment the background of elliptical epithelial cells to identify regions of interest [93]. Second, the unsupervised blemish separation method can accurately segment rough glandular borders based on the connections between the lumen and epithelial nuclei established by the least inertial axis and chain [94].

We summarize according to pathological image segmentation methods in different cancers. Table 1 organizes cancer types, Staining, applications, datasets, methods and corresponding references for each article used in this paper. In the past three years, DL frameworks and some mainstream segmentation models (such as U-Net, SegNet, etc.) are still important methods for segmenting histopathology images. Secondly, Table 2 organizes the commonly used datasets in Table 1 for your reference. Finally, table 2 also adds some datasets not mentioned in this paper.

Table 1 Overview of histopathological image segmentation methods for different cancers.

| Cancer type | Staining | Application | Dataset | Method | Ref |
|---|---|---|---|---|---|
| Colon | H&E | Segmentation | Warwick-QU(165WSIs) | Segnet | [45] |
| Colon | H&E | Segmentation | Warwick-UHCW(75WSIs) and Warwick-Osaka(50WSIs) | Persistent homology profiles (PHPs) + CNN | [47] |
| Colon | H&E | Classification & Segmentation | AiCOLO(396 WSIs), CRC-5000(625 WSIs), NCT-CRC-HE-100k(86 WSIs) and Warwick(16 WSIs) | UNet and SegNet | [48] |
| Colon | H&E | Classification & Segmentation | Kumar(30 WSIs), CoNSeP(41WSIs), CPM-15, CPM-17(32 WSIs), TNBC(50 WSIs), CRCHisto(100WSIs) | Hover-Net | [49] |
| Colon | H&E | Segmentation | Private(36WSIs) | Active contours mechanism | [50] |
| Colon | H&E | Segmentation | Warwick-QU(165WSIs) | Based on DenseNet to train DCIA+ConvCRFs | [51] |
| Colon | H&E | Segmentation | CoCaHis(82WSIs) | SVM, kNN, U-Net, U-Net++ and deeplabv3 classifiers | [52] |
| Colon | | Classification & | Private(68WSIs)+ | K-means algorithm, Haralick | [53] |

| | | Segmentation | Glas(165WSIs), | | |
|---|---|---|---|---|---|
| Colon | H&E | Segmentation | Gland Segmentation (GlaS) Challenge(775WSIs) | Preprocessing Techniques for Colon Histopathology Images | [54] |
| Colon | H&E | Segmentation | GlaS(775WSIs), Colorectal adenocarcinoma gland(CRAG)(214WSIs) | MILD-Net, MILD-Net+ | [55] |
| Colon | H&E | Segmentation | ADP database | HistoSegNet | [56] |
| Colon | H&E | Segmentation | GlaS(165WSIs), CRAG(173WSIs) | Novel label rectification method ECLR and a semi-supervised segmentation framework ECGSSL, a collaborative multi-task discriminative error prediction network DEP-Net | [57] |
| Colon | H&E | Segmentation | GlaS(165WSIs), CRAG(173WSIs) | Attention-Guided Deep Atrous-Residual U-Net | [58] |
| Colon | H&E | Segmentation | CoNIC(4981images) | SegCenterNet2 | [59] |
| Colon | H&E | Segmentation | Private (36WSIs) | Multilevel thresholding method | [60] |
| Breast | H&E | Segmentation | ICIAR2018 Breast Cancer dataset(400 images) | Mask RegionalConvolutional Network (Mask-RCNN) | [61] |
| Breast | H&E | Segmentation | DRYAD (51WSIs) | K-means clustering and watershed algorithms | [62] |
| Breast | H&E | Segmentation | BreakHis(7909WSIs) | PSO-Based Clustering Technique | [63] |
| Breast | H&E | Segmentation | UCSB(10WSIs) | Kapur, Minimum Cross Entropy, and Tsallis+Stochastic Fractal Search (SFS) algorithm | [64] |
| Breast | H&E | Segmentation | Andrew Janowczyk image database (143WSIs) | AlexNet | [65] |
| Breast | H&E | Segmentation | TNBC(50WSIs), Kidney dataset(730WSIs) | An effective encoder–decoder architecture with a separable convolution pyramid pooling network (SCPP-Net) | [66] |
| Breast | H&E | Segmentation | Public(141WSIs) | Deep residual neural network (DeepRNNetSeg) | [67] |
| Breast | H&E | Segmentation | BACH(440 WSIs), TCGA(8,897 cases) | ResNet50/U-Net | [68] |
| Breast | H&E | Segmentation | TNBC(32WSIs) | Deep Multi-Magnification Network (DMMN) | [69] |
| Breast | H&E | Segmentation | | CNN initialized active contour model with adaptive ellipse fitting (CoNNACaeF) | [70] |
| Prostate | H&E | Segmentation | Private(10WSIs) | Combining MR-based human segmentations with a semi-automated thresholding approach based on 18F-choline PET | [72] |
| Prostate | H&E | Segmentation | Private(367WSIs) | 3D watershed algorithm | [73] |
| Prostate | H&E | Segmentation | Andrew Janowczyk image database (143WSIs) | U-Net CNN with subsequent morphological filtering | [74] |
| Prostate | H&E | Segmentation | 2018 Data Science Bowl Grand Challenges(735WSIs) | Recurrent Residual CNNs based U-Net (R2U-Net) | [75] |
| Prostate | H&E and IHC | Segmentation | Private 1(102WSIs) | CNN | [76] |
| Prostate | H&E | Segmentation | | CNN including FCN-8s, two SegNet variants, and multi-scale U-Net | [78] |
| Prostate | H&E | Segmentation | Private(182WSIs) | SegNet , U-Net CNN | [79] |
| Prostate | H&E | Segmentation | Private(47WSIs) | Residual and Multi-resolution U-NET | [80] |

| Prostate | H&E | Segmentation | Private(486WSIs) | Image analysis techniques based on mathematical morphology | [81] |
|---|---|---|---|---|---|
| Prostate | H&E | Segmentation | Private(178WSIs) | Simple morphometric techniques+CNN | [82] |
| Prostate | H&E | Segmentation | Private(20WSIs) | A dense multi-path decoder + CNN | [83] |
| Prostate | H&E | Segmentation | TCGA(389WSIs) | NASNetLarge | [84] |
| Prostate | H&E | Segmentation | Private 2(1500 images) | A new hybrid DL method , a softmax-driven active contour model | [85] |
| Prostate | H&E | Segmentation | Private(40-50WSIs) | Gland Context Network (GCN) | [86] |
| Prostate | H&E | Segmentation | MICCAI 2015 Gland Segmentation Challenge Contest(16WSIs) | A unified two-parallel-branch deep neural network | [87] |
| Prostate | H&E | Segmentation | Dataset obtained from the Department of Pathology at Cedars-Sinai Medical Center(513 tiles + 30 WSIs) | An expectation maximization(EM)-based semi-supervised DL approach, Multi-scale U-Net | [88] |
| Stomach | H&E | Segmentation | Private(94WSIs), | A novel multi-task based DL model for HIstoPathOlOgy (named Deep-Hipo) that takes multi-scale patches simultaneously for accurate histopathological image analysis. | [90] |
| Other | H&E | Segmentation | TCGA(6481WSIs), kidney (2334), bronchus and lung (2168) and breast (1979) WSIs locations. | Weakly supervised ResNet50 architecture | [91] |
| liver | H&E | Segmentation | KMC liver dataset(80 images) | NucleiSegNet + a joint loss function | [92] |
| Colon | H&E | Segmentation | Private(10-12WSIs), Bilkent(72WSIs), Warwick-QU(165WSIs) | Rényi's two-dimensional entropy with a cultural algorithm (2DR $e$ CA) | [93] |
| Colon | H&E | Segmentation | MICCAI 2015 Gland Segmentation Challenge Contest(16WSIs) | Unsupervised taint separation method+U-Net | [94] |

Table 2 About different cancer histology image segmentation datasets and corresponding websites.

| | |
|---|---|
| Warwick-QU | https://warwick.ac.uk/fac/cross_fac/tia/data/glascontest/download |
| CRAG | https://warwick.ac.uk/fac/sci/dcs/research/tia/data/mildnet |
| CoNSep | https://warwick.ac.uk/fac/sci/dcs/research/tia/data/ |
| TNBC | https://zenodo.org/record/1174343#.YoHzechBxXK |
| Kumar | https://nucleisegmentationbenchmark.weebly.com/ |
| ADP database | https://drive.google.com/file/d/1jG1ojQKmvGjjjrRhCkaH0FDWM61tSgjL/view |
| DRYAD dataset | http://dx.doi.org/10.5061/dryad.pv85m |
| BreakHis | http://web.inf.ufpr.br/vri/breast-cancer-database |
| Warwick-UHCW | https://data.gov.uk/dataset/5ec4ae2d-b6af-4a71-a6cc-1fc48d5d317b/https-www-uhcw-nhs-uk-download-clientfiles-files-government-20transparency-20data-20dec-202017-csv |
| UCSB | http://dough.ece.ucsb.edu |
| Warwick-Osaka | https://warwick.ac.uk/TIA |

| TCGA | https://www.cancer.gov/about-nci/organization/ccg/research/structural-genomics/tcga/using-tcga/types |
|---|---|
| BACH | https://zenodo.org/record/3632035#.Yn0HpshBxXI |
| TNBC | https://rgcb.res.in/tnbcdb/download.php |
| Kidney dataset | https://archive.ics.uci.edu/ml/datasets/chronic_kidney_disease |
| KMC liver dataset | https://drive.google.com/drive/u/2/folders/1_lLVLKIqkpQa2YBC_76RUOXLLtdDIBoE |
| MoNuSAC2018 | https://monuseg.grand-challenge.org/Data/ |
| MoNuSAC2020 | https://monusac-2020.grand-challenge.org/Data/ |
| AGGC22 | https://aggc22.grand-challenge.org/Data/ |
| Private 1 | https://doi.org/10.5281/zenodo.1485967 |
| Private 2 | https://data.mendeley.com/datasets/h8bdwrtnr5/1 |

### 4.1.2. Mitosis detection

Mitosis is an important biomarker for cancer prognosis. DL is the primary method for detecting mitosis [95], [96]. Mitotic counts were assessed by counting mitoses in regions with the highest proliferative activity (visual selection) [97]. Hand-crafted feature convolutional neural networks (CNNs) can detect and classify mitosis end-to-end [98]. The deep neural network-based RetinaNet method provides fully annotated WSIs for mitotic maps, enabling the evaluation of mitotic detection algorithms and region-of-interest detection algorithms on the complete WSI [99]. The data augmentation strategy of convolutional neural networks combined with H and E staining significantly improves the robustness of detecting mitotic features [100]. Fig. 6 shows some detected mitosis examples. The mitotic figures identified by the hand-crafted CNN on the ICPR12 dataset are true positive (TP), false negative (FN) and false positive (FP) [101]. An efficient A supervised model of a DL framework has been demonstrated to detect mitotic features of breast histopathology in WSI images. The proposed architecture has improved to 92% precision, 88% recall and 90% F-1 score [95]. Mask RCNN, a multi-task DL framework for object detection and instance segmentation, can automatically detect mitosis from histological breast cancer slides [101]. Second, PartMitosis, a novel partially supervised framework based on two parallel deep fully convolutional networks, achieves F-scores of 0.575 and 0.698 on the 2014 ICPR dataset and AMIDA13 dataset, respectively, outperforming all previous mitosis detection systems [102]. A multi-stage mitotic cell detection method based on Faster Regional Convolutional Neural Network (Faster R-CNN) and deep CNN achieved F-scores of 0.858 and 0.691 on ICPR 2012 dataset and ICPR 2014 dataset [103]. Weakly supervised mitosis detection trains a segmentation model on a labelled dataset with centroid pixel format using a deep fully convolutional network and filters out low-confidence mitotic data on the segmentation map generated by the segmentation model [104]. Its model achieves 0.562 F-score and 0.673 F-score on ICPR 2014 MITOSIS dataset and AMIDA13 dataset.

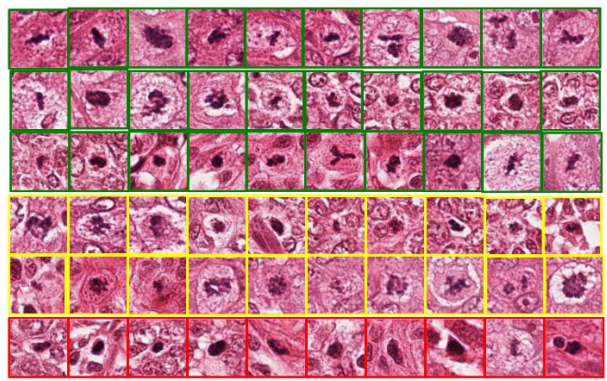

Fig. 6. The mitoses identified by HC+CNN on the ICPR12 dataset are TP (green rectangles), FN (yellow rectangles),

and FP (red rectangles). TP samples have obvious intensity, shape and texture, while FN samples are less obvious in intensity and shape. FP maps are more visually similar to mitotic maps than FN maps [105].

Transfer learning is also an important method for the automatic detection of breast cancer histopathology images. Among them, transfer learning-based deep CNN combined with Hybrid-CNN can be used to segment and detect mitosis in breast cancer histopathology images [106]. Automatic mitosis detection in breast histopathology images using deep transfer learning based on convolutional neural networks [107]. Experiments show that the pre-trained Convolutional Neural Network model outperforms conventionally used detection systems and provides at least 15% improvement in F-score on other state-of-the-art techniques [107]. Articles are organized by cancer types, Staining, applications, datasets, methods and corresponding references. the past three years

Table 3 Overview of histopathological mitotic assays for different cancers.

| Cancer type | Staining | Application | Dataset | Method | Ref |
|---|---|---|---|---|---|
| Breast | H&E | Detection | ICPR 2014(2127 frames), ICPR 2012(50 frames), AMIDA-13(585 frames) and projects (MICO ANR TecSan) | DL model(five convolution layers, four max-pooling layers, four rectified linear units (ReLU), and two fully connected layers. ) | [95] |
| Breast | H&E | Detection | ICPR 2012(50 frames), ICPR 2014 (2127 frames) | DeepMitosis | [96] |
| Breast | H&E | Detection | Private(90 cases), TNBC(298 ) | CNN | [97] |
| Breast | H&E | Detection | ICPR 2012(50 frames), ICPR 2014(2127 frames) | Free-handcrafted-feature CNN(AlexNet and U-Net) | [98] |
| Skin | H&E | Detection | Private(32WSIs) | RetinaNet | [99] |
| Breast | H&E | Detection | ICPR 2012(50 frames) | New color augmentation technique + CNN | [100] |
| Breast | H&E | Detection | ICPR 2012(50 frames) | CNN | [105] |
| Breast | H&E | Detection | ICPR 2012, ICPR 2014(2127 frames) | MaskMitosis (ResNet+ FPN) | [101] |
| Breast | H&E | Detection | ICPR 2012(50WSIs), ICPR 2014(1696 HPF images), AMIDA13(606 HPF images) | PartMitosis(dense prediction ConvNet) | [102] |
| Breast | H&E | Detection | ICPR 2012(50 frames), ICPR 2014(2127 frames) | Faster R-CNN and Deep CNN | [103] |
| Breast | H&E | Detection | ICPR 2012(50 frames), ICPR 2014(2127 frames), AMIDA-13(606 HPF images) and TUPAC16(656 HPF images) | SegMitos + deep CNN | [104] |
| Breast | H&E | Detection | TUPAC16(656 HPF images) | A multi-phase deep CNN | [108] |
| Breast | H&E | Detection | ICPR 2014(2127 frames) | Deep transfer learning(A pre-trained CNN + RF) | [107] |

Table 4 Summary of commonly used datasets in mitosis.

| ICPR 2012 | http://ludo17.free.fr/mitos_2012/download.html |
|---|---|
| ICPR 2014 | https://mitos-atypia-14.grand-challenge.org/Dataset/ |
| AMIDA-13 | http://amida13.isi.uu.nl/ |
| TUPAC16 | https://tupac.grand-challenge.org/Dataset/ |

## 4.2. Pathology image classification

Methods for analysing histopathology images of breast cancer are essential for diagnosing and prognosis of breast cancer. DL techniques are currently the dominant classification method [109]–[119]. Hybrid CNN models such as convolutional neural networks (CNN) and CNN+SVM were used to classify breast cancer microscopy images on the ICIAR2018 dataset [120]–[122]. DL frameworks based on entropy-based and confidence-enhancing strategies can reduce annotation costs [123]. Hybrid deep neural networks take advantage of convolutional kernel recurrent neural networks to preserve short- and long-term null correlation between blocks, and richer fusion of multi-level features allows for adequate image fusion [124], [125]. The multi-channel convolutional neural network (MCCNN), developed by connecting each layer with artificial neurons inspired by the human brain system, uses three channel inputs to extract computational features from each channel and connect them for multiple classifications, with an average accuracy of 96.4% for the model [126]. The VGG16-based VGGN-16 framework evaluated the performance of different classifiers on canine breast cancer (CMT) histopathology images and human breast cancer datasets [127]. An improved approach based on the VGG19 neural network fine-tuning the last convolution resulted in a final classification accuracy of 0.876 ± 0.026 [128]. Given that WSI is too large, InceptionV3 first extracted patch-level image features. Then, it passed the patch-level predictions through an integrated fusion framework involving majority voting, gradient boosting machine (GBM) and logistic regression to obtain image-level predictions [129]. The fine-tuned classification models of ResNet V15 and ResNet V152 on tissue array (TMA) can identify malignant or benign subtypes of cancer with high sensitivity [130]–[133]. Combining three fine-tuned models based on a multi-scale convolutional neural network (EMS-Net) is more accurate than other combinations [134]. Deep ResNet institutions with convolutional blocks and attention modules can extract richer, more fine-grained features from pathology images [135], [136]. Xception models and SVM classifiers with a 'radial basis function' kernel achieve high accuracy at 40X, 100X, 200X and 400X levels of magnification [137]. A sequential framework using multi-layer deep features extracted from fine-tuned DenseNet yields better performance at the highest layer compared to DenseNet [138], [139]. CNNs based on residual learning obtained 84.34% accuracy and 90.49% F1 score on classifying breast cancer histopathology images [140]. Better classification performance of the model under data enhancement. A novel neural network designed SE-ResNet module based on the residual module and Squeeze-and-Excitation block optimises the parameters of the model parameters [141]. Second, the new learning rate does not require complex fine-tuning to obtain a high-precision classification model. Two consecutive convolutional neural networks are connected in series, with the first neural network performing pre-training first to extract local features and the second network acquiring global information about the input image [142]. Attention mechanisms are embedded in the parallel structure of CNNs and RNNs, replacing batch normalisation with a switchable normalisation method and using targeted dropout regularisation techniques instead of the traditional dropout [143]. Histopathology image classification based on deep feature fusion and augmented routing (FE-BkCapsNet) extracts both convolutional and capsule features and integrates semantic and spatial features into a new capsule to obtain more discriminative information [144].

In classifying histopathology images of breast cancer, migration learning has improved the utilisation of the model compared to DL. Several experiments have demonstrated that migration learning is better than deep feature extraction and SVM classification [145]. A fine-tuned pre-trained VGG16 model with a logistic regression classifier achieved better classification results [146]. The design of the GoogLeNet model draws on a hybrid convolutional neural network

(CNN) architecture, with migration learning and data enhancement strategies to reduce the limitations of the data for GoogleNet [147]. With a combination of DL, migration learning, and generative adversarial networks, fine-tuning of the VGG16 and VGG19 networks was used to extract well-distinguishable cancer features from histopathology images then fed into a neuronal network for classification [148]. Fine-tuning of Google's Inception-V3 and ResNet50 convolutional neural networks (CNNs) was used to extract image features based on pre-trained models to obtain training models [149], [150]. Migration learning based on AlexNet, GoogleNet and ResNet can classify images in multiple cell and cell nucleus configurations [149].

A weakly-supervised classification framework for multi-instance learning reduces the hassle of labelling all the data and yields better results than several state-of-the-art MIL methods [151]. A new hierarchical cell-to-tissue map (HACT) consists of a low-level cell map and a high-level tissue map capturing the hierarchy of the cellular morphological organisation, encoding the relative spatial distribution of cells relative to tissue distribution. hCAT-NeT maps the representation of HCAT to histopathology images [152]. A weakly-supervised graphical convolutional network-based network can perform class classification in tissue microarrays with an accuracy of $0.9637 \pm 0.0131$ [153]. A novel CoMHisP framework for fuzzy support vector machines with intra-class density information (FSVM-WD) extracts image features by optimising the block size and calculates the centre of mass for each pixel to extract feature vectors. Experimental evidence shows that staining normalisation and magnification improve the accuracy of the CoMhISp model [154]. Deep manifold preserving autoencoder learns discriminative features by preserving the structure of the input dataset from the popular learning view and minimising the reconstruction error in the kidney-void learning view from a large amount of unlabelled data [155]. The scarcity of annotations has severely hampered the study of diagnostic models. Patch-based fully convolutional neural network (FCN) weakly supervised learning efficiently extracts deep representative features and generates feature representations of the entire WSI using a strategy of searching for different context-aware block selections and feature aggregation [156]. A random forest (RF) classifier produces the final image results.

Colon cancer analysis is a complex and time-consuming task, and pathologists can get visible results directly from classifying of histopathology images. The inter-collaboration of artificial intelligence and computational pathology will boon for physicians and patients. DL is the dominant method for classifying colon cancer pathology images [157]–[159]. Pairwise constrained regularised deep convolutional neural networks optimise the performance of traditional CNNs and update the parameters of CNNs by merging newly annotated samples to update the data [160].CNN models such as SqueezeNet-v1.1, MobileNet-v2, ResNet-18 and DenseNet-201 have been shown to have improved accuracy on available publicly available data sets with improved accuracy [161]. A parallel combination of the VGG16 model and CapsNet obtained an accuracy of 0.982 on a classification task after tissue microarray (TMA) core extraction and colour enhancement [162]. A VGG-based modified U-shaped framework performing training and inference tasks will split the mask to generate classification probabilities for the likelihood of a WSI being malignant [163]. To use a smaller model, RCCNet was compared with five state-of-the-art CNN models on the conventional colon cancer histology dataset "CRCHistoPhenotypes" and obtained an accuracy of 80.61% and a weighted average F1 score of 0.7887 [46]. The unsupervised approach to DL quantifies significant sub-regions by extracting a set of features learned directly on the image data and uses these quantified distributions for image representation and classification.

Biopsy specimens and radical prostatectomy specimens were graded according to the 2014 ISUP five-level Gleason grading system (five-level GG system). Several colour preprocessing methods such as colour enhancement, colour constancy, colour deconvolution and colour transfer did not improve classification accuracy when combined with convolutional networks. However, some preprocessing methods improved classification accuracy slightly when used in conjunction with texture-based methods [164]. Region-based convolutional neural networks use an epithelial network head and a grading network head for multi-task prediction [165]. Deep learning models and seven machine learning methods were compared for Gleason score changes and classification performance in the pT2 phase. The deep model obtained the highest F1 scores with pT2 tumours (0.849) and Gls changes (0.574) [166], [167]. A multi-resolution, multi-instance learning model was developed for better Gleason grading. The model uses saliency images to detect suspicious regions and classifies cancers at higher magnification for selected clays and can be trained without fine annotation using WSI labels with a grading accuracy of 92.7% [168], [169]. Retraining Inception-v3 automatically detects the use of Gleason patterns and identifies grade groups [170]. A weakly-supervised GCN-based approach to rank classification in tissue microarrays (TMAs) outperformed a self-supervised approach [153]. This is mainly because the model models cellular spatial organisation as a graph, better capturing tumour cells' value-added and community structure. A new set of complete and statistical local binary pattern (CSLBP) descriptors allows for Gleason grading of prostate cancer from H&E-stained WSI pathology images [171].

The classification of histological images of lung cancer is essential for determining the tumour grade and treatment of patients. However, subjective criteria influence the heterogeneity and assessment of lung cancer; therefore, an objective method for grading lung cancer is needed. Ten ML models and VGG16 models were trained on histological images of 867 adenocarcinoma (ADC) patients and 552 squamous cell carcinoma (SCC) patients and yielded the best classification models [172], [173]. CNNs and soft polling act as a pipeline for decision functions to identify solids, micropapillae, blisters, screen growth patterns, and non-tumour regions [174]. Deep learning models can automatically classify histological images of WSI [175]–[177]. The deep learning-based Convpath method is applied to regions of interest (ROI) of tumours in parallel through multiple threads and automatically classifies different types of nuclei in less time [178]. Second, CNNs based on the EfficientNet-B3 architecture use migration learning and weakly supervised learning strategies to train on training data from 3554 WSIs and detect cancers in WSIs [179]. Inception v3 can also classify histopathological images of non-small cell lung cancer and predict the ten most common mutated genes in lung adenocarcinoma (LUAD) from pre-trained models [180]. GAN is used as a strategy for data augmentation to help DCNN improve the accuracy of histological image identification [181].

We have summarised the pathological image classification methods based on their use in different cancers. Table 4 collates each article by cancer type, Staining, application, dataset, method and the corresponding references. In the last three years, DL frameworks and some mainstream classification models (e.g. VGG, ResNet, InceptionV3, DensetNet, etc.) have remained important methods for classifying histopathology images. Secondly, Table 5 collates the commonly used datasets in Table 4 for your reference use. Some datasets not mentioned in this paper have also been added to Table 5.

Table 4 Summary of the histological image classification of the different cancers.

| Cancer type | Staining | Application | Dataset | Method | Ref |
|---|---|---|---|---|---|
| Breast | H&E | Classification | ICIAR 2018 Grand Challenge(400 images), BISQUE(58 WSIs) | CNN(ResNeXt50) | [109] |
| Breast | H&E | Classification | ICIAR 2018(400 images) | Ensemble support vector machine (E-SVM), DCNNs (e.g., DenseNet-121, ResNet-50, multi-level InceptionV3, and multi-level VGG-16), dual-network orthogonal low-rank learning (DOLL) | [110] |
| Breast | H&E | Classification | BreakHis(7909WSIs) | CNN、LSTM, CNN+LSTM, Softmax and Support Vector Machine (SVM) layers have been used for the decision-making stage. | [111] |
| Breast | H&E | Classification | ICIAR 2018(400 images) | CNN(ResNet-50, InceptionV3 and VGG-16) | [112] |
| Breast | H&E | Classification | Private(554 WSIs) | CNN(VGG16, VGG19) | [113] |

| Breast | H&E | Classification | Bioimaging 2015(249 images) | K-means clustering algorithm, ImageNet, ResNet50-512, ResNet50-128, ResNet50, | [114] |
|---|---|---|---|---|---|
| Breast | H&E | Classification | ICIAR 2018 (400 images) | CNN(AlexNet、Inception-Net、ResNet) | [115] |
| Breast | H&E | Classification | BreaKHis(7909WSIs) | Transformer learning(Inception_V3, Inception_ResNet_V2) | [132] |
| Breast | H&E | Classification | Public (277524 images), BreaKHis(7909WSIs) | CNN | [117] |
| Breast | H&E | Classification | BreaKHis(7909WSIs) | (a) Convolutional Neural Network Raw Image (CNN-I); (b) Convolutional Neural Network CT Histogram (CNN-CH); (c) Convolutional Neural Network CT LBP (CNN-CL); (d) Convolutional Neural Network Discrete Fourier Transform (CNN-DF); (e) Convolutional Neural Network Discrete Cosine Transform (CNN-DC). | [118] |
| Breast | H&E | Classification | Bioimaging 2015, BreaKHis | DCNN(Inception)+ boosting trees classifier | [119] |
| Breast | H&E | Classification | ICIAR2018(400 images) | CNN(ResNet, Inception v3) + SVM | [120] |
| Breast | H&E | Classification | BreaKHis(7909WSIs), BACH | CNN + Squeeze-Excitation-Pruning (SEP), Hybrid CNN architecture (Inception module, residual network and Batch Normalization (BN)) | [121] |
| Breast | H&E | Classification | Privare(5000 images) | An accurate, reliable and active (ARA) image classification framework and introduce a new Bayesian Convolutional Neural Network (ARA-CNN) | [122] |
| Breast | H&E | Classification | BreaKHis(7909WSIs) | Deep active learning framework: An entropy-based strategy and a confidence-boosting strategy. | [123] |
| Breast | H&E | Classification | Private 1(3771 images) | Patch-wise CNN | [124] |
| Breast | H&E | Classification | Private 2(3771 images) | A Hybrid Convolutional and Recurrent Deep Neural Network | [125] |
| Prostate | H&E | Classification | Private(6000images) | The Multichannel Convolution Neural Network (MCCNN) | [126] |
| Prostate | H&E | Classification | CMTHis(352WSIs), BreakHis(7909WSIs) | VGGNet-16 | [127] |
| Breast | H&E | Classification | ICIAR 2018 (400 images) | A Dual Path Network (DPN) + gradient boosting machine (GBM), logistic regression(LR), and support vector machine (SVM) | [129] |
| Breast | H&E | Classification | BreaKHis(7909WSIs) | Inception_V3 and Inception_ResNet_V2 | [130] |
| Breast | H&E | Classification | ICIAR 2018 (400 images) | Fusing the output of two residual neural networks (ResNet) | [131] |
| Breast | H&E | Classification | BreaKHis(7909WSIs) | CNN(ResNet50)+LR, CNN+SVM | [132] |
| Breast | H&E | Classification | ICIAR 2018 (400 images) | ResNet | [133] |
| Breast | H&E | Classification | ICIAR 2018 (400 images) | Ensemble of MultiScale Networks (EMS-Net) | [134] |
| Breast | H&E | Classification | BreaKHis(7909WSIs) | A deep ResNet structure with Convolutional Block Attention Module (CBAM) | [135] |
| Breast | H&E | Classification | BreaKHis(7909WSIs) | ResNet-50 | [136] |
| Breast | H&E | Classification | BreaKHis(7909WSIs) | Xception + SVM | [137] |
| Breast | H&E | Classification | BreaKHis(7909WSIs) | Sequential framework which utilizes multi-layered deep features that are | [138] |

| | | | | extracted from fine-tuned DenseNet. | |
|---|---|---|---|---|---|
| Breast | H&E | Classification | BreakHis(7909WSIs) | Interleaved DenseNet with SENet (IDSNet) | [139] |
| Breast | H&E | Classification | BreakHis(7909WSIs) | A residual learning-based 152-layered CNN(ResHist) | [140] |
| Breast | H&E | Classification | BreakHis(7909WSIs) | CNN with small SE-ResNet module | [141] |
| Breast | H&E | Classification | ICIAR 2018 (400 images) | Two-Stage(image-wise + patch-wise) CNN | [142] |
| Breast | H&E | Classification | ICIAR 2018(400 images), Bioimaging2015(249 images), Extended Bioimaging2015(1319 images) | Parallel Structure Deep Neural Network Using CNN and RNN with an Attention Mechanism | [143] |
| Breast | H&E | Classification | BreakHis(7909WSIs) | Based on deep feature fusion and enhanced routing (FE-BkCapsNet) | [144] |
| Breast | H&E | Classification | BreakHis(7909WSIs) | Transformer learning(AlexNet and Vgg16) + SVM | [145] |
| Breast | H&E | Classification | BreakHis(7909WSIs) | Transformer learning(VGG16, VGG19, and ResNet50) | [146] |
| Breast | H&E | Classification | ICIAR 2018 (400 images) | CNN(GoogLeNet) | [147] |
| Breast | H&E | Classification | BreakHis(7909WSIs) | Transfer Learning and GAN | [148] |
| Breast | H&E | Classification | ICIAR 2018 (400 images) | Transfer Learning Using a Pre-trained Inception Resnet V2 | [149, p. 2] |
| Breast | H&E | Classification | BreakHis(7909WSIs) | Inception-v3 + SVM | [150] |
| Breast | H&E | Classification | BreakHis(7909WSIs) | Multiple Instance Learning (MIL) methods(APR, Diverse Density, MI-SVM, citation-kNN), MIL-CNN | [151] |
| Breast | H&E | Classification | BRACS(106WSIs) | GNN(HACT-Net) | [152] |
| Prostate | H&E | Classification | Private(886 images) | A weakly-supervised approach for grade classification in tissue micro-arrays (TMA) using graph convolutional networks (GCNs) | [153] |
| Breast | H&E | Classification | CMTHis(352 images) | A novel CoMHisP framework based on a fuzzy support vector machine with within-class density information (FSVM-WD) | [154] |
| Lung | H&E | Classification | SUCC Dataset(939 WSIs), TCGA | Weakly supervised + FCN, context-aware block + RF | [156] |
| Colon | H&E | Classification | Private(740 images) | The CADx DL-based AI model | [157] |
| Colon | H&E | Classification | Private(2513 images) | DL model | [158] |
| Colon | H&E | Classification | LC25000(images) | Image Sharpening Using Unsharp Masking, 2D Fourier Features and 2D Wavelet Features + CNN | [159] |
| Colon | H&E | Classification | CRCHistoPhenotypes (100 WSIs) | DNN | [160] |
| Colon | H&E | Classification | Warwick-QU(165WSIs), Epistroma(1376 WSIs), BreaKHis(7909 images), multi-class Kather(5000 images) | SqueezeNet-v1.1、MobileNet-v2、ResNet-18 and DenseNet-201 | [161] |
| Colon | H&E | Classification | Private(54WSIs) | Soft Voting Ensemble of one VGG and one CapsNet models | [162] |

| Colon | H&E | Classification | DigestPath 2019(750WSIs) | The framework contains an improved U-shape network with a VGG net as backbone, and two schemes for training and inference, respectively (the training scheme and inference scheme) | [163] |
|---|---|---|---|---|---|
| Colon | H&E | Classification | Agios Pavlos(300 images), BreakHis(7909WSIs), Cedars-Sinai(625 images), HICL(109 subjects), Kather Multiclass(5000 WSIs), Lymphoma(372 WSIs), Netherlands Cancer Institute(1295 wsiS), Vancouver General Hospital, Warwick-QU(165WSIs) | Color Pre-Processing(Color Augmentation, Color Deconvolution, Colour Normalization) +Image Descriptors(Hand-Designed Methods, Pre-Trained Convolutional Networks) + Further Pre-Processing Step | [164] |
| Prostate | H&E | Classification | Private(513 images) | A new region-based convolutional neural network framework for multi-task prediction using an epithelial network head and a grading network head. | [165] |
| Prostate | H&E | Classification | Private(1833WSIs) | DL model | [167] |
| Prostate | H&E | Classification | Private (20229WSIs) | A multi-resolution MIL-based (MRMIL) model | [168] |
| Prostate | H&E | Classification | Private(96WSIs) | CNN(Inception-v3) | [170] |
| Prostate | H&E | Classification | TCGA(312 WSIs) | A set of novel completed and statistical local binary pattern (CSLBP) descriptors | [171] |
| Lung | H&E | Classification | TCGA(78WSIs) | Image tiles augmentation + CNN training | [174] |
| Lung | H&E | Classification | Private(279WSIs) | CNN(ResNet) | [175] |
| Lung | H&E | Classification, Segmentation | NLST(208 WSIs), TCGA-LUAD(431 WSIs) | Mask-RCNN model | [176] |
| Lung | H&E | Classification | Private | DL model | [177] |
| Lung | H&E | Classification | NLST(208 WSIs), TCGA-LUAD(431 WSIs), Private(130 WSIs) | Convpath + DL model | [178] |
| Lung | H&E | Classification | TCGA-LUAD and TCGA-LUSC(608 WSIs), Private(3704 WSIs), CPTAC-LSCC(500 WSIs) | CNN(EfficientNet-B3) | [179] |
| Lung | H&E | Classification | NCI Genomic Data Commons(1634 WSIs) | CNN(inception v3) | [180] |
| Lung | H&E | Classification | Private(511 images) | GAN + DCNN | [181] |

Table 5 Data sets and corresponding websites on the classification of different cancer histology images.

| | |
|---|---|
| BISQUE | http://bioimage.ucsb.edu/research/bio-segmentation |
| Bioimaging 2015 | https://rdm.inesctec.pt/dataset/nis-2017-003 |
| BreaKHis | http://web.inf.ufpr.br/vri/breast-cancer-database |
| Public 1 | http://andrewjanowczyk.com/wp-static/IDC_regular_ps50_idx5.zip |
| NLST | https://biometry.nci.nih.gov/cdas/nlst/ |
| TCGA-LUAD | https://wiki.cancerimagingarchive.net/display/Public/TCGA-LUAD |
| TCGA-LUSC | https://www.cancerimagingarchive.net/ |
| ICIAR2018 | https://iciar2018-challenge.grand-challenge.org/Dataset/ |
| Private 1 | http://ear.ict.ac.cn/?page_id=1616 |
| Private 2 | http://ear.ict.ac.cn/?page_id=1576 |
| Camelyon 2016 | https://camelyon16.grand-challenge.org/Data/ |
| Camelyon 2017 | https://camelyon17.grand-challenge.org/ |
| BRACS | https://www.bracs.icar.cnr.it/download/ |
| LC25000 | https://academictorrents.com/details/7a638ed187a6180fd6e464b3666a6ea0499af4af |
| Epistroma | http://fimm.webmicroscope.net/supplements/epistroma |
| DigestPath 2019 | https://digestpath2019.grand-challenge.org/Dataset/ |
| CPTAC-LSCC | https://wiki.cancerimagingarchive.net/display/Public/CPTAC-LSCC |
| NLST | https://cdas.cancer.gov/datasets/nlst/ |

## 4.3. Quantification

The tumour-to-stromal ratio (TSR) is an independent prognostic factor for colon cancer (CRC) and other solid malignancies. Machine learning approaches have accelerated the clinical implementation of TSR. the ML framework is capable of nuanced survival prediction in patients with stage II CRC, outperforms the current state of the art and shows great potential in computational pathology and healthcare [182]. In stratified Cox multivariate analysis, the reproducibility and robustness of Immunoscore's first strong prognostic factor associated with the TNM classification system favours its implementation as a new component in cancer classification, called TNM-Immune [183]. Semi-automatic segmentation of all relevant tissues in rectal cancer histology by DL is applied to sensitive regions provided by experts. The visual and automated TSR method assigns patients to either the 'high stroma' or 'low stroma' group and then compares prognosis in terms of colon cancer specificity and disease-free survival time [184]. A method for predicting clinically relevant molecular phenotypes from WSI histopathology images based on DL models trained on 5700 samples can yield interpretable image features (HIFs). The method was used to output 6.7 HIFs and to quantify specific and biologically relevant features for five cancer types [185]. A convolutional neural network under a migration learning strategy segmented the HE patches of the WSI and subsequently outputs the TSR [186]. The TSR was then found to have the ability to significantly improve prognosis when combined with other risk factors in a Cox model [187].

The characterisation of the tumour microenvironment including extracellular vesicles (EV) is an important direction for the analysis of cancer. EV density increased with higher histological grading and shorter tumour-to-edge distances in breast cancer patients [188]. Portable multimodal nonlinear optical imaging systems not only provide real-time visualisation of the tumour microenvironment, but also offer the potential to use EV as a marker-free biomarker for cancer diagnosis and prognosis [188]. Quantitative diffusion metrics such as mean diffusion rate (MD) and fractional anisotropy (FA) from extended tensor imaging (DTI) are associated with other factors in the histological prognosis of breast cancer patients [189]. Qualitative assessment of residual cancer burden index lymph nodes and percentage tumour cellularity (TC) composition in invasive or in situ tumour beds (TB) can be used to predict overall survival. Deep neural networks mimicking pathologists can automatically extract features from images and visualise the potential for improved TC assessment [190].

Machine learning-based image analysis software can be used in the quantification of histological images of tally ischaemic stroke thrombi and to assess the correlation between clot composition and the density of Huntsfield units on computed tomography (CT) scans [191], [192]. Machine learning algorithms to incorporate the most relevant tissue phenotypes resulted in an 88% improvement in the accuracy of tumour progression prediction, directly improving critical treatment decisions for prostate cancer patients [193]. Several experiments demonstrated that CNN models outperformed fibrosis scoring (PEFS) in the classification task of trichrome-stained histological images and could correlate several clinical phenotypes [194]. Secondly, machine learning-based methods for liver histological assessment improve on the disadvantages of manual histological assessment, accurately characterise disease severity and heterogeneity, and sensitively quantify treatment response to activity scores [195]. Finally, histopathology images are also correlated with matched genomic, transcriptomic and survival data to enhance prognosis based on histopathological subtyping and grading [196].

## 5. Challenges of machine learning in computational pathology applications

Based on the literature survey, we concluded that the challenges of machine learning in computational pathology mainly include the following: standardization and normalization of data, lack of data identification, hardware limitations, and lack of transparency and interpretability. They are somewhat related. as shown in Fig .7. Among them, the lack of transparency and interpretability of machine learning technology is the main problem of current application landing.

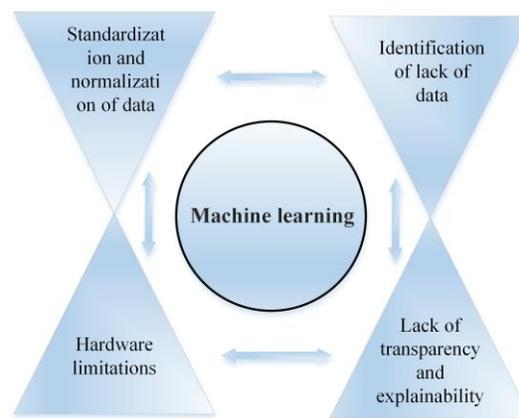

Fig .7 Challenges encountered by machine learning techniques in processing histopathology images.

**Challenge #1: Standardization and normalization of data**

The successful adaptation of WSI in digital pathology largely depends on every step of high-quality pathology slide preparation, including embedding, cutting, staining, and scanning [8]. Folded tissue sections during cutting, changes in staining and the presence of air bubbles during overlay slides, and different settings for brightness, intensity differences, average color, and border intensity during scanning can cause unreliable raw data and produce inaccurate results [197], [198]. Protocols and quality system controls need to be standardized to reduce noise in the data affecting the classification of slides. Different histopathological data require different algorithms for analysis, which requires a large amount of data to meet the needs of the model and make the algorithm more accurate. Standard data formats and canonical methods for data analysis should be adopted to combine datasets from different organizations and institutions and use them for training an algorithm to enhance the robustness of the model.

Different data sources can vary the classification accuracy. DICOM (www.dicomstandard.org) defines a medical image format that can be used for data exchange at a quality that meets clinical needs. A DICOM file contains a data header and image data, where the data header covers the patient's number, name, gender, and the number of frames and resolution of the image. The standard now provides the same support for WSI by processing large, tiled images into multiple frames and multiple images of different resolutions (http://dicom.nema.org/Dicom/DICOMWSI/)[8].

**Challenge #2: Lack of data identification**

Practice has proved that supervised learning models outperform models trained in other ways. Mainly, most supervised learning AI algorithms require many high-quality training images. Ideally, the images on which the model was directly trained are already labeled. This requires pathologists and specialists to hand-paint different (abnormal or malignant) regions in the image, distinguish between different cell types, and circle the cell outline and nuclear extent. Manual development consumes much time, labor cost and money cost. Crowdsourcing may be cheaper and faster, but may introduce noise. Second, the tedious manual labeling work can make doctors and experts impatient, and the subjective local field of view increases the error rate of image labeling. Currently, only a small number of labeled datasets are available for direct use by physicians. Therefore, there is an urgent need to develop active learning annotable models.

**Challenge #3: Hardware limitations**

The accuracy of diagnostic models in computational pathology mainly depends on a large amount of data, sufficient computing power, reliable hardware and a network environment [8]. WSI is a super-gigabit digital image. A standard WSI image has 50000x50000 pixels and a size of about 3GB. Therefore, larger image data requires a large amount of hardware storage space or cloud space to store the data. Although, at present, DL is still the main method in the field of computational pathology, especially when applying pathology graph analysis, a large number of high-efficiency image computing units can process data in parallel on GPU. The memory of the GPU and the memory of the host are also two factors that affect the training efficiency of image data. Therefore, high-performance graphics cards and large-capacity memory ensure WSI image computing power and shorten training time and test time. In addition, the sharing of WSI resources will increase the robustness of the algorithm. Therefore, whether it is local data transmission or cloud data transmission, the data bandwidth of the intranet and the Internet is also a factor that affects the quality of the model [8]. Only when these relevant conditions are met, computational pathology can continue to move forward to help doctors and patients solve more complex and multifaceted clinical problems and research tasks.

**Challenge #4: Lack of transparency and explainability**

The current interpretability of machine learning algorithms in computational pathologies, such as DL, is still low—the lack of interoperability of artificial neural networks as classifiers. DL is often thought of as a "black box," and while researchers should start working on innovative ways to interpret the results of models, these methods are still limited in their persuasiveness. There is currently no established way to easily explain why specific decisions are made by the network when processing histopathology scans [199]. Doctors and patients are still taking a wait-and-see approach to AI algorithms. In future research, we should focus on the innovation of the algorithm itself, and not blindly pursue the high accuracy of the model. The algorithm of transparency and interpretability is the main research direction at present. Attention mechanism, adversarial learning and deep reinforcement transfer learning can be used as a few algorithms to focus on. Finally, algorithms also need to be streamlined, portable, and trainable to help physicians learn more about these diagnostic models.

## 6. The promise of machine learning in computational pathology applications

Computational pathology has evolved to assist physicians in building clinical decision support tools to diagnose patients accurately. Extensive experiments have demonstrated that the technology can identify tumor cell images, count mitotic numbers, improve the accuracy of immunohistochemical scoring or apply standardized histological scoring criteria. This plays a key role in cancer screening and treatment [200]–[202]. DL-based diagnostic tools assess the space between immune cells within tumors and correlate with response to immunotherapy [203]. With the improvement of people's health awareness and the rapid development of health care technology, machine learning technology is gradually being integrated into our lives, such as health smart monitoring bracelets and watches, general practitioner robots, etc. 2019 is the first year of publication of the computational pathology white paper, and computational pathology has begun to increasingly involve multiple subspecialties such as lung, kidney, gastrointestinal, neurology, and gynecology. AI has made notable breakthroughs in specific tasks in these fields, such as image classification, segmentation, and quantification. Applying these algorithms in the future will liberate pathologists and improve the survival probability of cancer patients.

Growing medical big data, including genomics, proteomics, informatics, and WSI images, are expected to be integrated into multi-omics cancer data [8]. Machine learning algorithms with rich data will facilitate the intersection and fusion of digital pathology, molecular pathology and informatics pathology. This will accelerate the rapid development of AI-assisted computational pathology. The interconnection of global information, improved infrastructure, and enhanced computing power have provided convenience for the collection and processing of multi-omics data. In addition, the sharing mechanism of cloud data enables researchers to develop algorithms collaboratively.

## 7. Conclusion

In the era of rapid development of computer-assisted pathology, databases, cloud storage, and servers have gradually become an integral part of pathology's daily research and practice. In addition, the development of multiomics integrates genomics, transcriptomics, proteomics, metabolomics, pathomics, and radiomics and combines bioinformatics and machine learning algorithms to provide personalized precision medicine for patients from multiple perspectives. Therefore, computational pathology is a valuable method for disease diagnosis, prognosis and treatment. Furthermore, some doctors and pathologists have recognized the mature application of ML methods. In addition, Cross-country, cross-team, and cross-research collaborations will enrich multi-omics databases. Finally, computational pathology will have the potential to change and improve the problems facing the current healthcare system, although there are still many challenges and hurdles that remain unsolved.

## Acknowledgments.

This work was supported by National Key R&D Program of China 2017YFB0202602, 2018YFC0910405, 2017YFC1311003, 2016YFC1302500, 2016YFB0200400, 2017YFB0202104; NSFC Grants U19A2067, 61772543, U1435222, 61625202, 61272056; Science Foundation for Distinguished Young Scholars of Hunan Province (2020JJ2009); Science Foundation of Changsha kq2004010; JZ20195242029, JH20199142034，Z202069420652; The Funds of Peng Cheng Lab, State Key Laboratory of Chemo/Biosensing and Chemometrics; the Fundamental Research Funds for the Central Universities, and Guangdong Provincial Department of Science and Technology under grant No. 2016B090918122.